\documentclass[two column,superscriptaddress,groupedaddress amssymb, amsmath,nobibnotes,nofootinbib, aps, showpacs, prl]{revtex4}
\usepackage{color}
\usepackage[dvips]{graphicx}
\usepackage{graphicx}

\begin{document}
\title{Nodal Superconducting Gap Structure in the Quasi-One-Dimensional Cs$_2$Cr$_3$As$_3$ Investigated Using $\mu$µSR measurements}
\author{Devashibhai Adroja}
\email{devashibhai.adroja@stfc.ac.uk}
\affiliation{ISIS Facility, Rutherford Appleton Laboratory, Chilton, Didcot Oxon, OX11 0QX, United Kingdom} 
\affiliation{Highly Correlated Matter Research Group, Physics Department, University of Johannesburg, PO Box 524, Auckland Park 2006, South Africa}
\author{Amitava Bhattacharyya}
\affiliation{ISIS Facility, Rutherford Appleton Laboratory, Chilton, Didcot Oxon, OX11 0QX, United Kingdom} 
\affiliation{Highly Correlated Matter Research Group, Physics Department, University of Johannesburg, PO Box 524, Auckland Park 2006, South Africa}
\affiliation{Department of Physics, Ramakrishna Mission Vivekananda University, Howrah-711202, India}
\author{Michael Smidman}
\affiliation{Center for Correlated Matter and Department of Physics, Zhejiang University, Hangzhou 310058, China}
\author{Adrian Hillier}
\affiliation{ISIS Facility, Rutherford Appleton Laboratory, Chilton, Didcot Oxon, OX11 0QX, United Kingdom} 
 \author{Yu Feng}
\affiliation{State Key Laboratory of Surface Physics and Department of Physics, Fudan University, Shanghai 200433, China}
\author{Bingying Pan}
\affiliation{State Key Laboratory of Surface Physics and Department of Physics, Fudan University, Shanghai 200433, China}
\author{Jun Zhao}
\affiliation{State Key Laboratory of Surface Physics and Department of Physics, Fudan University, Shanghai 200433, China}
\author{Martin R. Lees}
\affiliation{Department of Physics, University of Warwick, Coventry, CV4 7AL, United Kingdom}
\author{Andre Strydom}
\affiliation{Highly Correlated Matter Research Group, Physics Department, University of Johannesburg, PO Box 524, Auckland Park 2006, South Africa}
\author{Pabitra K. Biswas}
\affiliation{ISIS Facility, Rutherford Appleton Laboratory, Chilton, Didcot Oxon, OX11 0QX, United Kingdom} 

\date{\today}

\begin{abstract}

The superconducting ground state of the newly discovered superconductor Cs$_2$Cr$_3$As$_3$ [$T_{\bf c}\sim$ 2.1(1) K] with a quasi-one-dimensional crystal structure has been investigated using magnetization  and muon-spin relaxation and rotation ($\mu$SR), both zero-field (ZF) and transverse-field (TF), measurements. Our ZF $\mu$SR measurements reveal the presence of spin fluctuations below 4 K and the ZF relaxation rate ($\lambda$) shows an enhancement below $T_{\bf c}\sim$ 2.1 K, which might indicate that the superconducting state is unconventional. This observation suggests that the  electrons are paired via unconventional channels such as spin fluctuations, as proposed on the basis of theoretical models. Our analysis of the TF $\mu$SR results shows that the temperature dependence of  the superfluid density is  fitted  better with a nodal gap structure than an isotropic  $s$-wave model for the superconducting gap.  The observation of a nodal gap in Cs$_2$Cr$_3$As$_3$  is consistent with that observed in the isostructural K$_2$Cr$_3$As$_3$ compound through TF $\mu$SR measurements.  Furthermore,  from our TF $\mu$SR study we have estimated the magnetic penetration depth of the polycrystalline sample  $\lambda_{\mathrm{L}}$$(0)$ = 954 nm, superconducting carrier density $n_s = 4.98 \times 10^{26}~ $m$^{-3}$, and carriers' effective-mass enhancement $m^*$ = 1.61\textit{m}$_{e}$.
\end{abstract}

\pacs{74.70.Xa, 74.25.Op, 75.40.Cx}

\maketitle
\section{1. Introduction}

In a conventional superconductor, the binding of electrons into the paired states, known as the Cooper pairs,  that collectively carry the supercurrent is mediated by lattice vibrations  or phonons, which is the fundamental  principle of the Bardeen-Cooper-Schrieffer (BCS) theory~\cite{B.C.S.}. The BCS theory often fails to describe the superconductivity (SC) observed in strongly correlated materials. Several strongly correlated superconducting materials, having magnetic  \textit{f}$-$ or \textit{d}$-$ electron  elements, exhibit unconventional SC and various theoretical models based on magnetic interactions (magnetic glue)   and spin fluctuations have been proposed to understand these superconductors~\cite{U.C.S.}. The superconducting gap structure of strongly correlated \textit{f}$-$ and \textit{d}$-$ electron superconductors is very important for understanding the physics of unconventional pairing mechanisms in these classes of materials. 

Unconventional superconductivity has been  observed in  high-temperature cuprates ~\cite{H.T.S.C.}, iron pnictides ~\cite{FeAs} and heavy fermion materials~\cite{H.F.S.C.}, which  have strong electronic correlations and quasi-two-dimensionality. It is of great interest to explore possible unconventional SC in a quasi-one-dimensional (Q1D) material with significant electron correlations. The recently discovered superconductors with a Q1D crystal structure, K$_2$Cr$_3$As$_3$ \textit{T}$_{\bf c}\sim$ 6.1 K, Rb$_2$Cr$_3$As$_3$ \textit{T}$_{\bf c}\sim$ 4.8 K and Cs$_2$Cr$_3$As$_3$ \textit{T}$_{\bf c}\sim$ 2.2 K have been intensively investigated both experimentally and theoretically~\cite{J. Bao,T. Kong, Z. Tang, Z. Tang1, G. M. Pang, DTA, X. Wu, H. Z., G.M. Pang-2, Li-Da Zhang, Chao Cao, Matt} as they are strong candidates for a multiband triplet pairing state as well as spin fluctuation mediated superconductivity from the \textit{d}-electrons of the Cr ions. In recent years the search for triplet superconductivity has been one of the major research efforts partly due to its intrinsic connection to topologically related physics and quantum computation. These new superconductors are conjectured to possess an unconventional pairing mechanism~\cite{J. Bao, Z. Tang1,G. M. Pang,DTA, X. Wu,H. Z.}. There are several pieces of experimental evidence for this. Firstly, the upper critical field $H_{c2}$ perpendicular to Cr-chain is significantly larger than the Pauli limit, but parallel to the Cr-chain it exhibits  paramagnetically limited behavior, indicating that the BCS-type pairing is unfavorable ~\cite{T. Kong,CAM, Balakirev,X.F. Wang}. Secondly, strong electronic correlations which are a common feature of unconventional superconductivity were revealed by a large electronic specific heat coefficient and non-Fermi liquid transport behavior ~\cite{J. Bao,X.F. Wang}. This is consistent with the Q1D crystalline structure of $A_2$Cr$_3$As$_3$ ($A$ = K, Rb and Cs) and represents a possible realization of a Luttinger liquid state ~\cite{H. Z. Zhi}. Thirdly, line nodal gap symmetry was revealed by London penetration depth and superfluid density measurements of K$_2$Cr$_3$As$_3$ ~\cite{G. M. Pang, DTA}. Fourthly, the effect  of nonmagnetic impurities reveals that $\textit{T}_{\bf c}$ decreases significantly for K$_2$Cr$_3$As$_3$ ~\cite{Y.Liu}, in accordance with the generalized Abrikosov-Gor’kov pair-breaking theory~\cite{AGP}, which supports  non-\textit{s}-wave superconductivity.

\begin{figure}[t]
\centering
\includegraphics[trim= 0cm 0cm 0cm 0cm, clip=true, totalheight=0.35\textheight, angle=0]{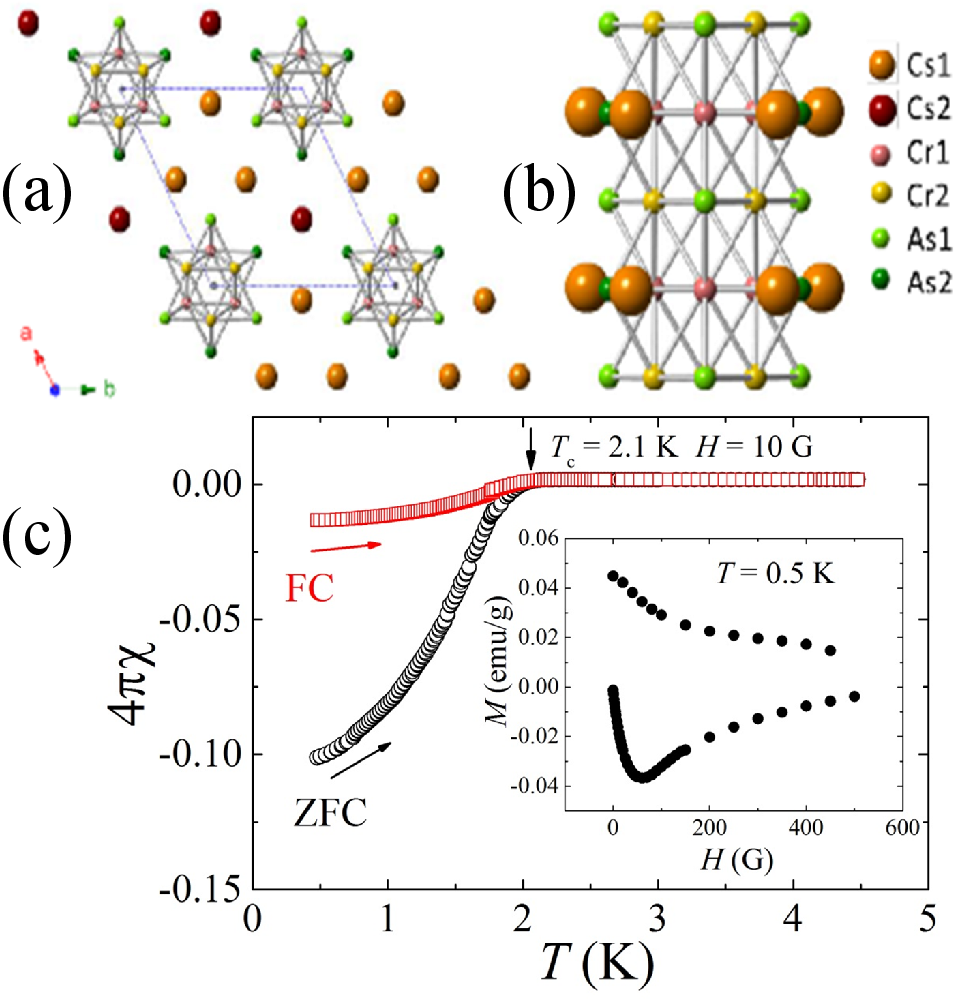}

\caption {(Color online) (a) The quasi-1D crystal structure of Cs$_2$Cr$_3$As$_3$ and (b) the $c$-axis view of the crystal structure showing the $\left[\mathrm{Cr}_3\mathrm{As}_3\right]_{\infty}$ chains. (c) The temperature dependence of the dc magnetic susceptibility measured in the zero-field cooled state (ZFC) and field cooled state (FC) of Cs$_2$Cr$_3$As$_3$ in the presence of an applied magnetic field of 10~G. The inset in (c) shows the magnetization versus field at 0.5 K.}
\end{figure}

Theoretically, by using density functional theory (DFT) calculations, X. Wu {\it et al.} predicted K$_2$Cr$_3$As$_3$ to be near a novel in-out co-planar magnetically ordered state and possess strong spin fluctuations~\cite{X. Wu,H. Z.}. Furthermore, it has been shown that a minimum three-band model based on the $d_{z^2}$ , $d_{xy}$ and $d_{x^2-y^2}$ orbitals of one Cr sublattice can capture the band structure near the Fermi surfaces. First principle calculations for K$_2$Cr$_3$As$_3$ reveal that one three-dimensional Fermi surface and two quasi-one-dimensional Fermi sheets cross the Fermi energy~\cite{H.Jiang, P.Alemany, A.Subedi}. In both the weak and strong coupling limits, the standard random phase approximation (RPA) and mean-field solutions consistently yield a triplet $p_z-$wave pairing as the leading pairing symmetry for physically realistic parameters~\cite{X. Wu, X. Wu2, W. Wu3,H. Zhong, Y. Zhou}. The triplet pairing is driven by ferromagnetic fluctuations within the sublattice~\cite{X. Wu,H. Z.,H.Jiang}. The gap function of the pairing state possesses line gap nodes on the $k_z$ = 0 plane of the Fermi surface. So it is highly likely that electrons are paired via unconventional channels such as spin fluctuations in $A_2$Cr$_3$As$_3$. Nuclear magnetic resonance (NMR) and nuclear quadrupole resonance (NQR)   measurements indeed reveal the enhancement of spin fluctuations approaching $\it{T}_{\bf c}$ in K$_2$Cr$_3$As$_3$~\cite{H. Z. Zhi} and  Rb$_2$Cr$_3$As$_3$~\cite{J. Yang}.  Furthermore  Y. Zhou {\it et al.}~\cite{Y. Zhou} have shown theoretically that at small Hubbard $U$ and moderate Hund's coupling, the pairing arises from the 3-dimensional (3D)  $\gamma$ band and has $f_{y(3x^2 -y^2)}$ symmetry, which gives line nodes in the gap function. At large $U$, a fully gapped $p$-wave state dominates on the quasi-1D $\alpha$-band. There are large numbers of experimental results as well as theoretical calculations reported for $A_2$Cr$_3$As$_3$ ($A$ = K and Rb), however, not much experimental work has been reported on Cs$_2$Cr$_3$As$_3$. This is due to the difficulties in synthesizing samples of this material,  because of its high air sensitivity~\cite{Z. Tang}. We have been able to synthesize a good quality powder sample of  Cs$_2$Cr$_3$As$_3$  and have investigated this material using magnetization and $\mu$SR measurements.

\section{2. Experimental Details}

A polycrystalline sample of Cs$_2$Cr$_3$As$_3$ was prepared as described by Tang {\it et al.}~\cite{Z. Tang}. This sample was characterized using x-ray diffraction and magnetic susceptibility. The magnetization data were measured using a Quantum Design Superconducting Quantum Interference Device magnetometer equipped with an iQuantum He-3 insert between 0.4 and 4.5 K.  Muon spin relaxation/rotation ($\mu$SR) experiments were carried out on the MUSR spectrometer at the ISIS pulsed muon source of the Rutherford Appleton Laboratory, UK~\cite{sll}. The $\mu$SR measurements were performed in zero$-$field (ZF) and transverse$-$field (TF) modes. A polycrystalline sample of Cs$_2$Cr$_3$As$_3$ was mounted in a sealed titanium (99.99\%) sample holder under  He-exchange gas, which was placed in a dilution refrigerator operating in the temperature range of 50 mK$-$5 K. It should be noted that we used small pieces of the sample (not fine powder) to minimize the decomposition of the sample, as the sample is very air sensitive.  Using an active compensation system the stray magnetic fields at the sample position were canceled to a level of 1~mG. TF$-\mu$SR experiments were performed in the superconducting mixed state in an applied field of 400 G, well above the lower critical field of $H_{c1}$= 10 G of this material.~Data were collected in the  (a) field$-$cooled (FC) mode, where the magnetic field was applied above the superconducting transition and the sample was then cooled down to base temperature and (b) zero field cooled (ZFC) mode, where first the sample was cooled down to 0.05 K in ZF and then the magnetic field was applied. Muon spin relaxation is a dynamic method that allows one to resolve the nature of the pairing symmetry in superconductors~\cite{js}. The mixed or vortex state in the case of type-II superconductors gives rise to a spatial distribution of local magnetic fields; which demonstrates itself in the $\mu$SR signal through a relaxation of the muon polarization. The asymmetry of the muon decay in ZF is calculated by, $G_z(t) =[ {N_F(t) -\alpha N_B(t)}]/[{N_F(t)+\alpha N_B(t)}]$, where $N_F(t)$ and $N_B(t)$ are the number of counts at the detectors in the forward and backward positions respectively and $\alpha$ is a constant determined from calibration measurements made in the paramagnetic state with a small (20~G) applied transverse magnetic field. The data were analyzed using the free software package WiMDA~\cite{FPW}.

\begin{figure}[t]
\vskip -0.0 cm
\centering
\includegraphics[width = 8 cm]{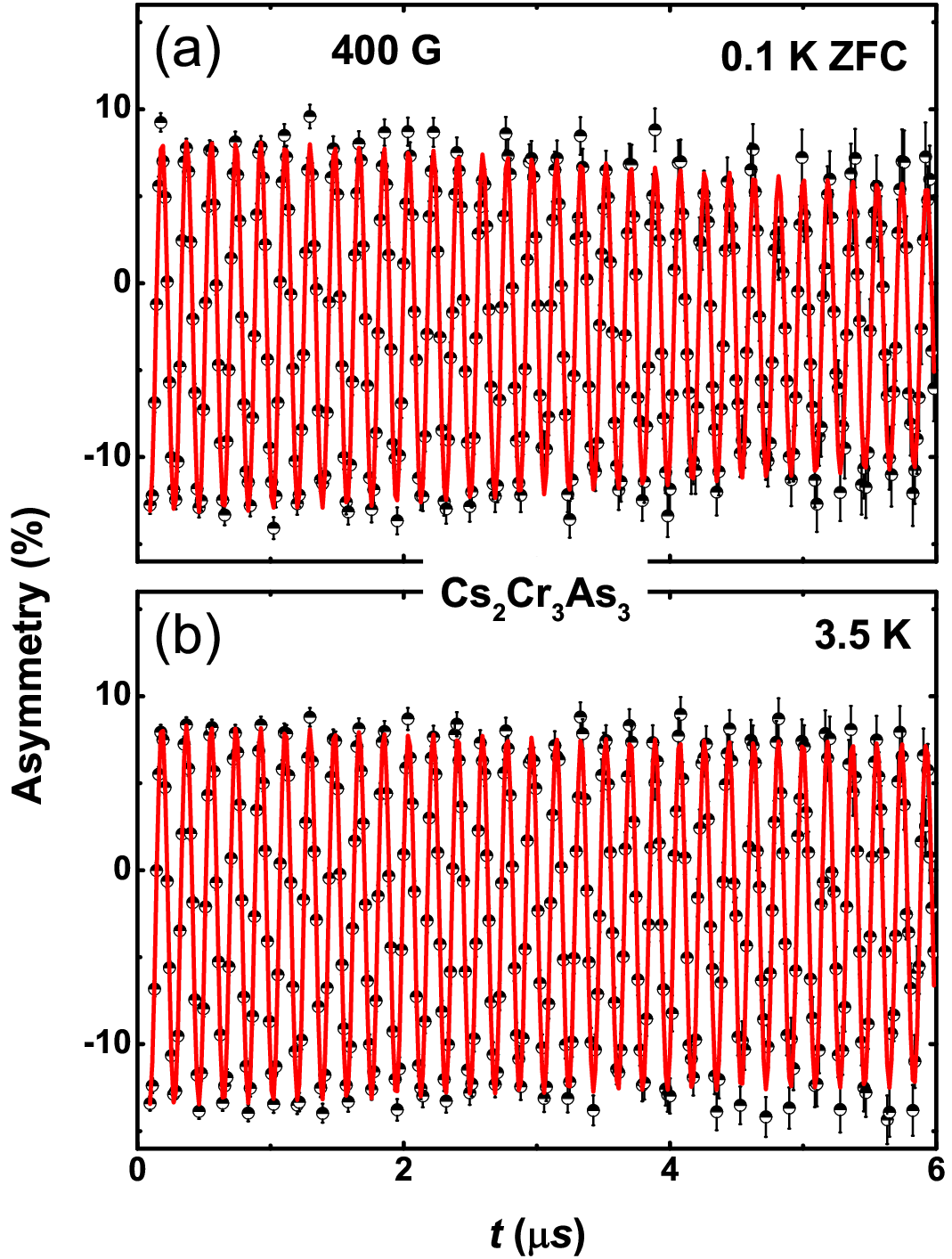}
\caption {(Color online) Transverse-field $\mu$SR time spectra (one component) for Cs$_2$Cr$_3$As$_3$ collected (a) at $T$ = 0.1 K and (b) at $T$ = 3.5 K in an applied magnetic field $H$ = 400 G for the zero-field cool (ZFC) state.}
\end{figure} 

\par
\section{3. Results and discussions}

\begin{figure}[t]
\centering
\includegraphics[width = 7cm]{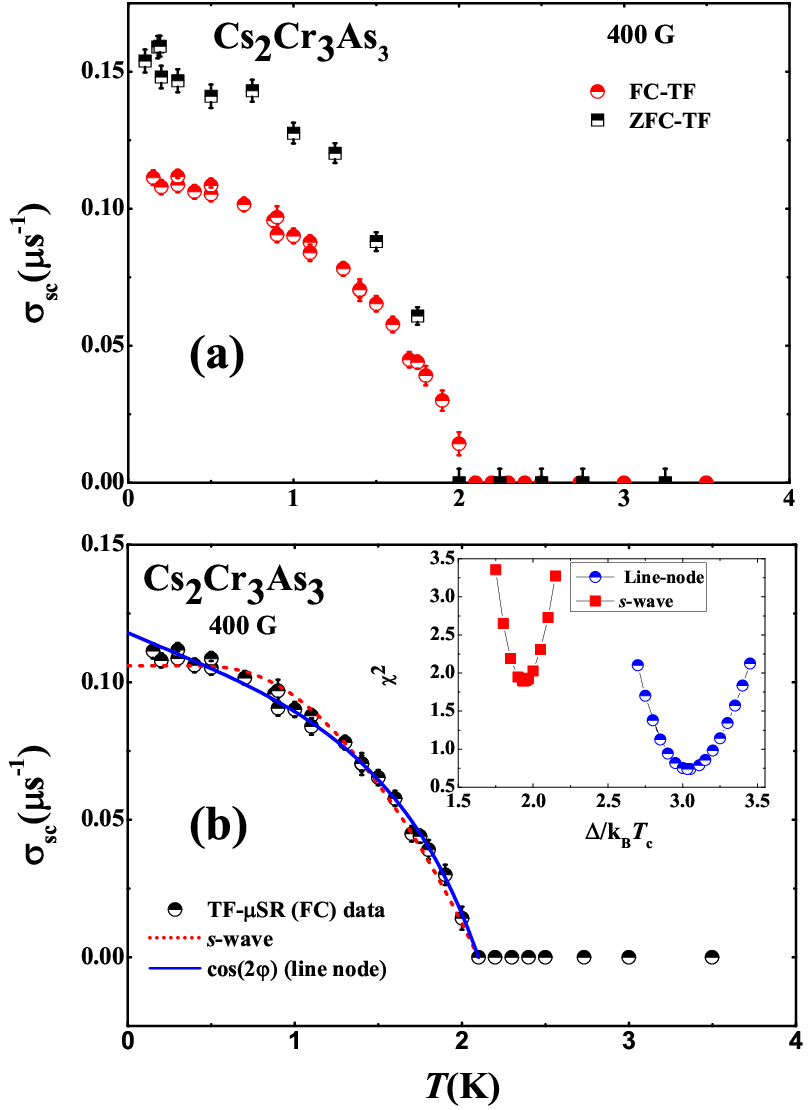}
\caption {(Color online) (a) Temperature dependence of  the muon depolarization rate $\sigma_{sc}(T)$ of Cs$_2$Cr$_3$As$_3$ collected in a magnetic field of 400 G in zero-field cooled (ZFC) and field cooled (FC) modes. (b) $\sigma_{sc}(T)$ of FC mode (symbols) and the  lines are the fits to the data using  Eq. 2. The short-dashed red line shows the fit using an isotropic single-gap $s$-wave model with $\Delta_0/k_{\mathrm{B}}T_{\mathrm{c}}$ = 1.94$\pm0.01$ and the solid blue line shows the fit to a nodal gap model with  $\Delta_0/k_{\mathrm{B}}T_{\mathrm{c}}=3.0\pm 0.2$, respectively. The inset shows the plot of quality of fit $\chi^2$ versus $\Delta_0/k_{\mathrm{B}}T_{\mathrm{c}}$.}
\end{figure} 

\begin{figure}[t]
\vskip -0 cm
\centering
\includegraphics[width = 7 cm]{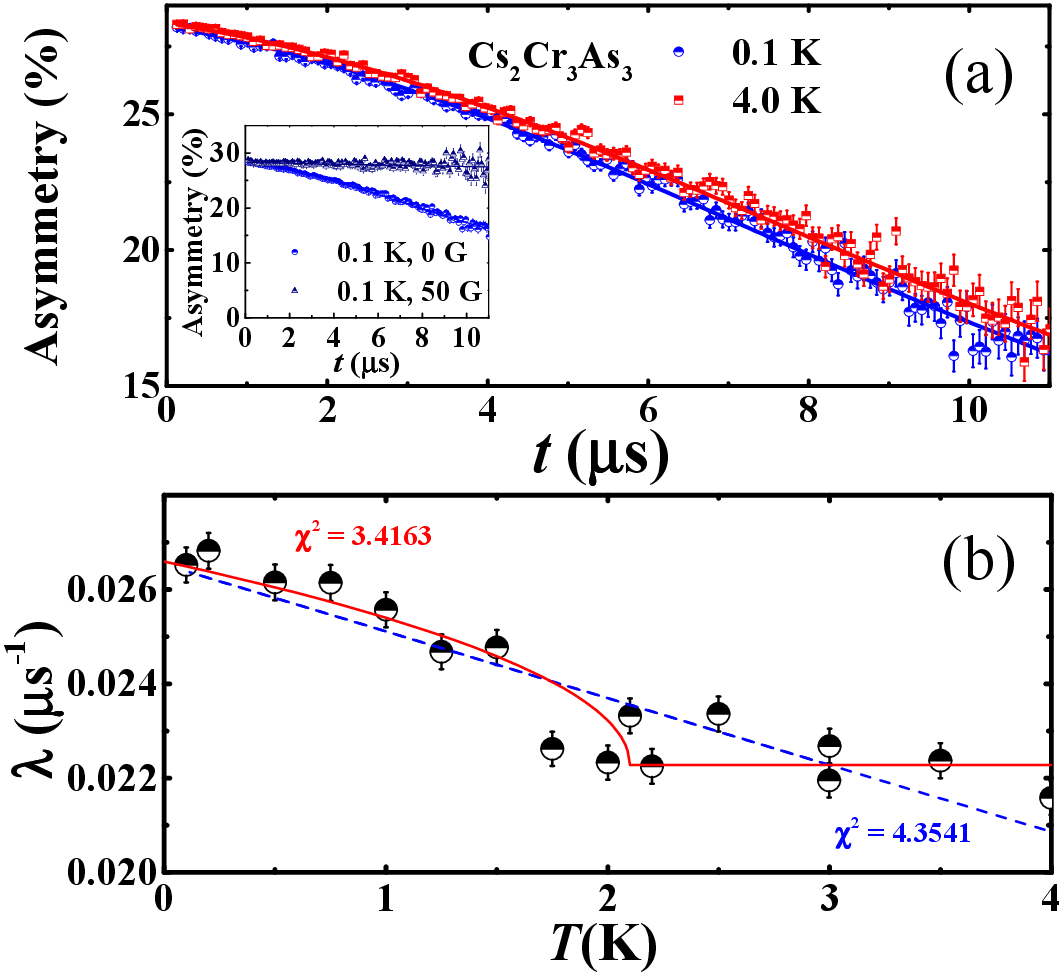}
\caption {(Color online) (a) Zero-field $\mu$SR time spectra for Cs$_2$Cr$_3$As$_3$ collected at 0.1 K ( circles) and 4.0 K (red squares) are shown together with lines that are least squares fits to the data using Eq. (5).  (b) The temperature dependence of the electronic relaxation rate measured in zero magnetic field of Cs$_2$Cr$_3$As$_3$, where  $T_{\bf c}$ = 2.1 K is shown by the vertical dotted line. The inset in (a) shows the spectra measured in 50 G longitudinal field together with ZF field results for comparison. In Fig.4b, the blue dotted  line shows the fit to a linear behavior ($\lambda(T)$=$a1$+$b1$*$\it{T}$) and the solid red line shows the fit to the BCS-type function ($\lambda(T)$=$a2$+$b2$*(1-$\it{T}$/$\it{T}$$_c$)$^{1/2}$ with $T_c$=2.1K fixed from the susceptibility}.
\end{figure}

The analysis of the powder x-ray diffraction at 300 K reveals that the sample is single phase  and crystallizes with space group $P\bar{6}m2$ (No. 187). The hexagonal crystal structure of Cs$_2$Cr$_3$As$_3$  is shown in Figs. 1(a-b). The Q1D feature of Cs$_2$Cr$_3$As$_3$ is manifested by the chains of [Cr$_3$As$_3$]$_{ \infty}$ octahedra (or double-walled subnanotubes) running along the $c$ direction (Fig.1b) which are separated by columns of Cs$^{+}$ ions, in contrast to the layered iron-pnictide and copper-oxide high $T_{\mathrm{c}}$  superconductors. Magnetic susceptibility measurements in an applied field of 10 G show that superconductivity occurs at 2.1(1) K and the superconducting volume fraction is  close to 10\% at 0.5 K  (Fig. 1(c)), indicating the bulk nature of superconductivity in  Cs$_2$Cr$_3$As$_3$. The small volume fraction of the superconductivity observed in our sample through the magnetic susceptibility measurements is in agreement with only 6\% superconducting volume fraction reported in the first report by  Tang {\it et al.} ~\cite{Z. Tang}. Another possible reason for a small SC volume fraction observed in our magnetic susceptibility measurements is that these measurements were carried out three months after our $\mu$SR study and it is likely that the sample partially decomposed during this time. The magnetization isotherm $M\left(H\right)$ curve at 0.5 K (inset of Fig.1(c)) shows a typical behavior for type-II superconductivity. We have estimated the lower critical field ($H_{c1}$) of 10 G from  the initial deviation from the linear behavior. The reported value of the upper critical field $H_{c2}$=64.5 kG~\cite{Z. Tang} is higher than the  Pauli limit, $\mu_{0}H_{P} = 18.4T_{\mathrm{c}} = 38.64$~ kG~\cite{CAM}, indicating unconventional superconductivity.  Using this value of $H_{c2}$=64.5 kG, we have estimated the superconducting coherence length $\xi = 7.14$~nm. In zero field, the temperature-dependent resistivity of Cs$_2$Cr$_3$As$_3$ is metallic~\cite{Z. Tang} and exhibits linear temperature dependence  between 50 K   and just above $T_{\mathrm{c}}$, indicating non-Fermi-liquid behavior and suggests the importance of spin fluctuations ~\cite{Z. Tang}.  At the superconducting transition, the dimensionless specific heat jump is  $\Delta$C/$\gamma$ $T_{\mathrm{c}}$ =0.4, which is smaller than the simple $\it {s}-$wave BCS prediction 1.43~\cite{Z. Tang}. The smaller value of the observed jump in the heat capacity was attributed to degradation of the sample~\cite{Z. Tang}.

\par

Figures 2 (a) and (b) show the TF$-\mu$SR precession signals above and below $T_{\bf c}$ obtained in ZFC mode with an applied field of 400 G (well above $H_{c1}\sim$ 10 G but below $H_{c2}\sim$ 64.5 kG). The observed decay of the $\mu$SR signal with time below $T_{\bf c}$ is due to the inhomogeneous field distribution of the flux-line lattice. We have used an oscillatory decaying Gaussian function to fit the TF$-\mu$SR asymmetry spectra, which is given below,

\begin{equation}
\begin{split}
G_{z1}(t) = A_1\rm{cos}(2\pi \nu_1 t+\phi_1)\rm{exp}\left({\frac{-\sigma^2t^2}{2}}\right)\\ + A_2\rm{cos}(2\pi \nu_2 t+\phi_2),
\end{split}
\end{equation}

\noindent where $\nu_1$ and $\nu_2$ are the frequencies of the muon precession signal from the superconducting fraction of the sample and from the background signal from the Ti-sample holder, respectively, $\phi_i$ ($i$ = 1, 2) are the initial phase offsets.   $A_1$ and $A_2$ are the muon initial asymmetries associated with the sample and background, respectively. The fits reveal the relative values of $A_1=35\%$ and $A_2=65\%$. It is noted that the value of $A_2$ is higher than $A_1$. One possible reason for this is that the muons have to first pass through the titanium foil (30 $\mu$m thick) and then stop in the sample. Another reason for the large  background asymmetry is due to the fact that we have used small pieces of the sample rather than fine powder to minimize the decomposition of the sample and hence it is possible that pieces might have settled down at the bottom of the sample holder when mounted vertically on the instrument. This also results in a higher fraction of the muons stopping directly in the titanium sample holder. Furthermore, the sample is very air sensitive and some parts of the sample may have decomposed, which also contributes to $A_2$. It is to be noted that our attempt to fit three components, one for the superconducting part of the sample, a second term  for the non-superconducting part of the sample and a third term for the sample holder, was not successful. This may be because the decay associated with muons stopped in the normal region is likely to only arise from nuclear moments and therefore it may be difficult to distinguish between muons stopped in the normal regions and muons stopped in the holder when fitting the data below $\it {T}_{\bf c}$. In that sense, the large background asymmetry $A_2$ = 0.65 will incorporate both these components. On the other hand, the muons stopped in the superconducting regions have a significantly higher decay rate. As a result, we are confident that the component with $A_1$ = 0.35 corresponds to muons stopped in the superconducting regions and that the decay rate from this component, from which we obtain the penetration depth, is not influenced by the presence of any non-superconducting regions. Following the explanation given above, we used only two components (see Eq.~1) in our TF-data analysis.  In Eq.~1 the first term contains the total relaxation rate $\sigma$ from the superconducting fraction of the sample; there are contributions from the vortex lattice ($\sigma_{sc}$) and nuclear dipole moments   ($\sigma_{nm}$), which is assumed to be constant over the entire temperature range below $T_{\bf c}$ [ where $\sigma$ = $\sqrt{(\sigma_{sc}^2+\sigma_{nm}^2)}$]. The contribution from the vortex lattice, $\sigma_{sc}$, was determined by quadratically subtracting the background nuclear dipolar relaxation rate obtained from the spectra measured above $\it {T}_{\bf c}$. As $\sigma_{sc}$ is directly related to the superfluid density, it can be modeled by \cite{Prozorov}

\begin{equation}
\frac{\sigma_{sc}(T)}{\sigma_{sc}(0)} = 1 + 2 \left\langle\int_{\Delta_k}^{\infty}\frac{\partial f}{\partial E}\frac{E{\rm d}E}{\sqrt{E^2-\Delta_k^2}}\right\rangle_{\rm FS},
\label{RhoS}
\end{equation}

\noindent where $f=\left[1+\exp\left(-E/k_{\mathrm{B}}T\right)\right]^{-1}$ is the Fermi function and the brackets correspond to an average over the Fermi surface. The gap is given by $\Delta(T, \varphi)$=$\Delta_0 \delta(T/\it {T}_c)g(\varphi)$, where $\varphi$ is the azimuthal angle along the Fermi surface. At present, no reliable experimental value of $\Delta C/\gamma \it {T}_c$ is available for Cs$_2$Cr$_3$As$_3$,  we therefore have used the BCS  formula for the temperature dependence  of the gap, which is given by $\delta(T/T_c)$ =tanh$[(1.82){(1.018(\it {T}_c/T-1))}^{0.51}]$~\cite{Prozorov, UBe13}, with g($\varphi$)~=~1 for the $s$-wave model  and g($\varphi$)=$|\cos(2\varphi)|$  for the $d$-wave model with line nodes ~\cite{Prozorov, UBe13}.

\par
Figure 3 (a) shows the temperature dependence of $\sigma_{sc}$, measured in an applied field of 400 G collected in two different modes: zero-field-cooled (ZFC) and  field cooled (FC). The temperature dependence of  $\sigma_{sc}$ increases with decreasing temperature confirming the presence of a flux-line lattice and indicates a decrease of the magnetic penetration depth with decreasing temperature. Comparing the ZFC and FC data reveals a substantial difference. In the ZFC mode, $\sigma_{sc}$ increases with decreasing temperature faster than for the FC data and thus points to differences in the number of the pinning sites, and trapping energies which are altered by the magnetic field history of the sample. From the analysis of the observed temperature dependence of $\sigma_{sc}$, using different models, the nature of the superconducting gap can be determined. We have analyzed the temperature dependence of  $\sigma_{sc}$ based on two models: an isotropic s-wave gap model  and a line nodes model.  The fit to the $\sigma_{sc}(T)$ data of Cs$_2$Cr$_3$As$_3$ by a single isotropic $\it{s}$-wave  gap  using Eq.(2) gives  $\Delta_0/k_{\mathrm{B}}T_{\mathrm{c}} = 1.94\pm0.01$ and $\sigma_{sc}(0)=0.106\pm0.001~\mu$s$^{-1}$ with the goodness of the fit  $\chi^2=1.94$ (see Fig.~3b, red short-dashed line).  The fit to the nodal model ( solid line Fig.~3b) shows better agreement than that of $\it{s}$-wave model and gives a larger value of $\Delta_0$/$k_B$$\it{T}_{\bf c}$ = 3.0$\pm$0.2 and $\sigma_{sc}(0)=0.118\pm0.002~\mu$s$^{-1}$ with $\chi^2=0.75$. We also tried $\it{s}$-wave and  nodal $\it{d}$-wave fits with various fixed values of $\Delta_0$/$k_B$$T_{\bf c}$ and allowing only $\sigma_{sc}(0)$  to vary, to compare the  ($\chi^2$) values between these two models (see the inset in Fig.~3b).  These plots of  $\chi^2$ confirm that our data fit better to the nodal gap model than the isotropic gap $s$-wave model. Therefore our $\mu$SR analysis is more consistent with Cs$_2$Cr$_3$As$_3$ having line nodes than being fully gapped. This is in agreement with the tunnel diode oscillator and $\mu$SR results of K$_2$Cr$_3$As$_3$~\cite{G. M. Pang, DTA}, which also support the presence of line nodes in the superconducting gap. 

Furthermore, the large value of $\Delta_0/k_{\mathrm{B}}T_{\mathrm{c}} = 3.0\pm0.2$ obtained from the line nodes $\it{d}$-wave fit indicates the presence of strong coupling and unconventional superconductivity, which is in line with that observed in K$_2$Cr$_3$As$_3$ of $\Delta_0/k_{\mathrm{B}}T_{\mathrm{c}} = 3.2\pm0.1$~\cite {DTA}. In addition, a $^{75}$As NMR study on K$_2$Cr$_3$As$_3$ and Rb$_2$Cr$_3$As$_3$ reveals the absence of a Hebel-Slichter coherence peak of $1/T_1$ just below $T_{\bf c}$, which is followed by a steep decrease, in analogy with unconventional superconductors in higher dimensions with point or line nodes in the energy gap ~\cite{H. Z. Zhi, J. Yang}. Furthermore, the  $T^5$ variation of the spin-lattice relaxation rate $1/T_1$ at low temperatures in Rb$_2$Cr$_3$As$_3$ suggests unconventional superconductivity with point nodes in the gap function ~\cite{J. Yang}. Both the NMR Knight Shift and $1/T_1T$ increase upon cooling in the normal state below 100 K, which are consistent with ferromagnetic spin fluctuations in both K and Rb samples~\cite{H. Z. Zhi, J. Yang}.

The Hebel-Slichter coherence peak of $1/T_1$ is a crucial test for the validity of the description of the superconducting state based on the conventional isotropic BCS s-wave model. The absence of the coherence peak in $1/T_1$ of $A_2$Cr$_3$As$_3$ ($A$ = K and Rb) suggests that an isotropic s-wave model is not an appropriate model to explain the gap symmetry.  These results along with our  $\mu$SR analysis of K$_2$Cr$_3$As$_3$ and Cs$_2$Cr$_3$As$_3$ suggest the presence of a nodal gap in all three compounds.

The  muon spin depolarization rate ($\sigma_{sc}$) below $T_{\bf c}$  is related to the magnetic penetration depth ($\lambda$). For a triangular lattice,~\cite{jes,amato,chia} 

\begin{equation}
\frac{\sigma_{sc}(T)^2}{\gamma_\mu^2}= \frac{0.00371\phi_0^2}{\lambda^4(T)}
\end{equation}

 where $\gamma_\mu/2\pi$ = 135.5 MHz/T is the muon gyromagnetic ratio and $\phi_0$ = 2.07$\times$10$^{-15}$ T m$^2$ is the flux quantum. This relation between $\sigma_{sc}$ and $\lambda$ is valid for 0.13/$\kappa^{2}$$<<$(H/H$_{c2}$)$<<$1, where $\kappa$=$\lambda$/$\xi$$\gg$70~\cite{Brandt}. As with other phenomenological parameters characterizing a superconducting state, the penetration depth can also be related to microscopic quantities. Using London theory~\cite{jes}, $\lambda_L^2= m^{*}c^2/4\pi n_s e^2$, where $m^* = (1+\lambda_{e-ph})m_e$ is the effective mass and $n_s$ is the density of superconducting carriers. Within this simple picture, $\lambda_L$ is independent of magnetic field. $\lambda_{e-ph}$ is the electron-phonon coupling constant, which can be estimated from $\Theta_D$ and $T_{\mathrm{c}}$ using McMillan's relation~\cite{mcm} 

\begin{equation}
\lambda_{e-ph}=\frac{1.04+\mu^*\ln(\Theta_D/1.45T_{\bf c})}{(1-0.62\mu^*)\ln(\Theta_D/1.45T_{\bf c})+1.04}
\end{equation}

where $\mu^*$ is the repulsive screened Coulomb parameter usually assigned as $\mu^*$ = 0.13.   For Cs$_2$Cr$_3$As$_3$  we have used $T_{\bf c}$ = 2.1 K and $\Theta_D$ = 160 K , which together with $\mu^*$ = 0.13, we have estimated $\lambda_{e-ph}$ = 0.75. Further assuming that roughly all the normal state carriers ($n_e$) contribute to the superconductivity (i.e., $n_s\approx n_e$),   we have estimated the magnetic penetration depth $\lambda$, superconducting carrier density $n_s$, and effective-mass enhancement $m^*$  to be $\lambda_L(0)$ = 954(9) nm (from the nodal fit), $n_s$ = 4.98$\times$10$^{26}$ carriers/m$^3$, and $m^*$ = 1.61 $m_e$, respectively.  It should be noted that the estimated value of $n_s$ represents the lower bound considering the stability of the sample when exposed to air. More details on these calculations can be found in Refs.~\cite{adsd,vkasd,dtasd}. Very similar values of $n_s$ and $m^*$  were estimated for K$_2$Cr$_3$As$_3$, but the $\lambda_L(0)$= 432 nm was a factor of 2.2 smaller ~\cite{DTA}.

\par
\
The  measured ZF$-\mu$SR spectra of Cs$_2$Cr$_3$As$_3$ are shown in Fig.~4(a) for $T$ = 0.1 and 4.0 K. In ZF relaxation experiments, any muons stopping in the titanium sample holder give a time independent background. The absence of a muon precession signal in the spectra in Fig. 4(a), rules out the presence of a long-range magnetic  ordered ground state in this compound. One possibility is that the muon$-$spin relaxation is due to either static, randomly oriented local fields associated with the nuclear moments at the muon site or the fluctuating electronic moments. The ZF$-\mu$SR data are well described by,

\begin{equation}
G_{z2}(t) =A_1 G_{KT}(t)e^{-\lambda t}+A_{bg}
\end{equation}
where 
\begin{equation}
G_{KT}(t)=[\frac{1}{3}+\frac{2}{3}(1-\sigma_{KT}^2t^2)exp(-\frac{\sigma_{KT}^2t^2}{2})]
\end{equation}

\noindent where the Gaussian Kubo-Toyabe $G_{KT}(t)$~\cite{KT}  functional form is expected from an isotropic
Gaussian distribution of randomly oriented static (or quasistatic) local fields at the muon sites. $\lambda$ is the electronic relaxation rate, $A_1$ is the initial asymmetry and $A_{bg}$ is the background.  The parameters $A_1$, and $A_{bg}$ are found to be temperature independent. The value of $\sigma$${_{KT}}$=0.078~$\mu$$s^{-1}$ was estimated from the 4 K data and was kept fixed in our analysis. It is interesting to note that $\lambda$ is finite at 4 K and increases [Fig. 4 (b)] with an onset temperature of $\sim2.1\pm0.1$~K, indicating the slowing down of electronic spin fluctuations correlated with the superconductivity.  We cannot rule out a weak, but temperature independent contribution to the relaxation rate from the non-superconducting part of the sample, but this does not alter our main conclusion. To check whether the $\lambda (T)$ shows an increase that is correlated with $\it{T}$$_c$, we have made two types of fit, (1)  A linear function across the whole temperature range ($\lambda(T)$=$a1$+$b1$*$\it{T}$) (see dotted blue line in  Fig.~4(b)) and (2) a BCS-type function ($\lambda(T)$=$a2$+$b2$*(1-$\it{T}$/$\it{T}$$_c$)$^{1/2}$ with $T_c$=2.1K fixed from the susceptibility (see solid red line in Fig.~4(b)). This analysis shows that the value of the goodness of fit (taking into account the number of free parameters) for option (1) is  $\chi^2$=4.3541, while for option (2) is   $\chi^2$=3.4163. This suggests that the $\lambda (T)$ shows an increase below $\it{T}$$_c$. This observation suggests that the superconductivity of Cs$_2$Cr$_3$As$_3$ is most probably associated with the spin fluctuation mechanism, which is supported by the NMR and NQR measurements on $A_2$Cr$_3$As$_3$ ($A$ = K and Rb)~\cite{H. Z. Zhi, J. Yang}. Increases in $\lambda$ at $\it{T}$$_c$ have been observed in the superconducting states of Sr$_2$RuO$_4$~\cite{gm}, LaNiC$_2$~\cite{ad1}, Lu$_5$Rh$_6$Sn$_{18}$~\cite{ab1} and Y$_5$Rh$_6$Sn$_{18}$~\cite{ab2}, but above $\it{T}$$_c$ $\lambda$ remains almost constant in these compounds.   This type of increase in $\lambda$  has been explained in terms of a signature of a coherent internal field with a very low frequency associated with time reversal symmetry (TRS) breaking  by Luke {\it et al.}~\cite{gm} for Sr$_2$RuO$_4$. Therefore based on these observations and considering the temperature dependence of $\lambda$ above and below $T_{\mathrm{c}}$ in Cs$_2$Cr$_3$As$_3$, we suggest that spin fluctuations play an important role in the superconductivity.  A very similar temperature dependent behavior of the ZF $\mu$SR depolarization rate is observed above and below the superconducting transition in $\it{R}$RuB$_{2}$ ($\it{R}$ = Lu, Y) by Barker {\it et al.}~\cite{J. Barker}, which has been attributed to the presence of quasistatic magnetic fluctuations. Finally we would like to mention the very recent NMR/NQR results on the normal state of Cs$_2$Cr$_3$As$_3$ (the sample does not show an onset of superconductivity down to 1.6K) ~\cite{hzhi-nmr}. The results show that a strong enhancement of Cr spin fluctuations (quasi-1D) at low temperatures is absent in their powder sample. They have mentioned that the underlying cause of this observation is not clear. Further they have mentioned that perhaps a naive quasi-1D picture, and hence the Tomonaga-Luttinger picture, is not valid in this material.

\section{4. Conclusions} 

In conclusion, we have presented zero-field (ZF) and transverse field (TF) muon spin rotation ($\mu$SR) measurements in the normal and  the superconducting state of  Cs$_2$Cr$_3$As$_3$, which has a quasi-one-dimensional crystal structure.  Our ZF $\mu$SR data reveal the presence of spin fluctuations at 4 K, which become stronger passing through the superconducting transition $T_{\mathrm{c}}=2.1$~K, indicating that the spin fluctuations are important for the superconducting state. The change of the ZF relaxation rate $\Delta$$\lambda$  from 4 K (i.e above $T_{\bf c}$) to the lowest temperature is $5 \times 10^{-3}\mu$s$^{-1}$ in Cs$_2$Cr$_3$As$_3$, which is a factor of 20 larger than that observed in K$_2$Cr$_3$As$_3$ ($\Delta\lambda = 2.5 \times 10^{-4}\mu$s$^{-1}$) despite the fact that the ratio of the $\it{T}_{\bf c}$'s is  0.34. Considering Cs is a larger ion than K we would expect the lattice expansion to lead to reduction in the spin fluctuations in Cs$_2$Cr$_3$As$_3$ when compared to K$_2$Cr$_3$As$_3$, which is not the case and hence remains to be understood. From the TF $\mu$SR we have determined the muon depolarization rate in ZFC and FC modes associated with the vortex-lattice.  The temperature dependence of $\sigma_{sc}$  fits  better to a nodal gap model than an isotropic $s-$wave model. Further, the value of $\Delta_0/k_{\mathrm{B}}T_{\mathrm{c}} = 3.0\pm0.2$ is obtained from the nodal gap model fit, indicating the presence of strong coupling and unconventional superconductivity in Cs$_2$Cr$_3$As$_3$. These results are in agreement with our previous findings for K$_2$Cr$_3$As$_3$. Considering the possibility of a multi-band nature of superconductivity in $A_2$Cr$_3$As$_3$ ($A$ = K, Rb and Cs), one would expect more complex behavior of the gap function and hence the conclusions obtained from our TF $\mu$SR study are in line with this.  Further confirmation of the presence of  line nodes in the superconducting gap requires $\mu$SR, heat capacity and thermal conductivity investigations of good quality single crystals of Cs$_2$Cr$_3$As$_3$.  

\
\
\section*{ACKNOWLEDGEMENT}

DTA and ADH would like to thank CMPC-STFC, grant number CMPC-09108, for financial support. DTA would like to thank JSPS for the award of their fellowship and Prof. Takabatake for his kind hospitality at Hiroshima University and the Royal Society of London for the UK-China Newton funding. AB would like to acknowledge DST for Inspire Faculty Research Grant, FRC of UJ, NRF of South Africa and ISIS-STFC for funding support.  Work at Fudan University was supported by the Ministry of Science and Technology of China (Program 973: 2015CB921302) and the National Natural Science Foundation of China (No. 91421106). 
AMS thanks the SA-NRF (Grant 93549) and UJ Research Committee for  financial support.

\end{document}